\documentclass[preprint]{aastex}
\usepackage{graphicx}

\begin{document}        
\title{DYNAMICS OF FE-NI BUBBLES IN YOUNG SUPERNOVA REMNANTS}
\author{John M. Blondin, Kazimierz J. Borkowski, and
Stephen P. Reynolds}
\affil{Department of Physics, North Carolina State         
University, Raleigh, NC 27695}

\begin{abstract}
Observations of core-collapse supernovae (SNe) have revealed the presence of 
extensive mixing of radioactive material in SN ejecta. The mixing of 
radioactive material, mostly freshly synthesized Ni, is not complete, which
leads to a two-phase SN ejecta structure. The low-density phase consists
of Fe bubbles, created by the energy input from radioactive Co and Ni, 
surrounded by compressed high-density metal-rich ejecta. 

We report on the theoretical investigation of supernova remnant (SNR) dynamics 
with the two-phase 
SN ejecta. We first present 3-dimensional hydrodynamic simulations of
a single Fe bubble immersed in an outer ejecta
envelope, and compare the results with previous work on shock-cloud 
interactions.  We then consider randomly distributed 
Fe bubbles with an average volume filling fraction of $1/2$.
We find that the presence of Fe bubbles 
leads to vigorous turbulence and mixing of Fe with other heavy elements and 
with the ambient normal-abundance gas. The turbulent energy can be an
order of magnitude larger than in the case of smooth ejecta. 
A significant fraction of the shocked 
ejecta is found in narrow filaments and clumps moving with radial velocities 
larger than the velocity of the forward shock.
Observational consequences of the
two-phase ejecta on SNR X-ray spectra and images are briefly mentioned.

\end{abstract}        
        
\section{INTRODUCTION}

The presence of large-scale mixing of ejecta in core-collapse supernovae 
(SNe) has been well established on both observational and theoretical
grounds. The early, unexpected emergence of X-rays and $\gamma$-rays shortly
after the explosion of SN 1987A provided  dramatic evidence for
mixing of radioactive Ni throughout the He core and H-rich
envelope. Extensive mixing of Ni is also required to explain the light curve 
and spectral observations of SN 1987A (see McCray 1993 for a review 
of SN 1987A). From studies of light curves and spectra of numerous 
core-collapse SNe, it is now known that such 
large-scale mixing is nearly always present. Another line of evidence is 
provided by studies of meteoritic graphite grains which condensed in SN 
ejecta (Travaglio et al. 1999 and references therein).  Such grains 
apparently contained radioactive, short-lived isotopes 
such as $^{44}$Ti at the time of their formation, which implies
extensive mixing of freshly-synthesized material deep in the SN ejecta with
the C-rich layer at the bottom of the He core. This overwhelming observational
evidence for mixing is consistent with our present understanding of the
collapse and explosion of massive stars. The neutrino-driven  
Rayleigh-Taylor instability just outside the proto-neutron star leads to
convective motions, strengthening the post-bounce shock and maybe even
making the explosion possible in many cases (e.~g., Mezzacappa et al. 1998;
Kifonidis et al. 2000). Further mixing occurs later
during the SN explosion because of Rayleigh-Taylor instabilities generated at
interfaces between stellar layers with different chemical composition
(e.g., Fryxell, Arnett, \& Mueller 1991; Hachisu et al. 1992; Kifonidis et al. 2000). 

Large-scale turbulence generated during the explosion appears not to be
sufficient to mix SN ejecta completely, but only macroscopically.  In
Cas A, optical observations revealed the presence of ejecta with very
different chemical abundances, such as O-, S-, Ar-, and Ca-rich ejecta
knots, which were macroscopically mixed during the SN explosion as
evidenced by the lack of spatial stratification expected in the
absence of mixing (e.~g., Fesen \& Gunderson 1996). The most recent IR and
X-ray observations of Cas A provide further evidence for this
macroscopic but not microscopic mixing (Arendt et al. 1999; Douvion et
al. 1999; Hughes et al. 2000; Hwang et al. 2000).  Spectroscopic observations 
of numerous
SNe also confirm the presence of clumpy ejecta (Spyromilio
1994). Detailed studies of SN 1987A lead to the same conclusion.

In SN 1987A, the inhomogeneities take a particularly interesting form:
Fe-Ni bubbles (Li et al. 1993), inferred from observations of Fe, Co,
and Ni lines.  Ni-rich ejecta clumps, transported outwards by
turbulent motions, are heated by their own radioactive energy input,
and expand in the ambient substrate of other heavy elements, forming
low-density Fe bubbles. The SN structure is then expected to be like a
Swiss cheese, with Fe bubbles occupying a substantial ($\sim 0.5$)
fraction of the ejecta volume (Li et al. 1993; Basko 1994).  These
bubbles should lead to vigorous turbulence and mixing in young supernova
remnants (SNRs) in
general.

Most of ejecta in young SNRs can be seen in X-rays, where 
modern X-ray
observatories such as {\it Chandra} and XMM-{\it Newton} have begun to
provide high-quality information. Interpretation of these observations
must be done in the framework of appropriate hydrodynamical models.
But realistic simulations for young SNRs, which include the presence of 
Fe bubbles in SN ejecta, are lacking. We report here on such 3-dimensional 
hydrodynamical simulations (preliminary results based on 2-D simulations were 
reported by Borkowski et al. 2000). 
Our goal is to study the basic hydrodynamics of the
interaction by investigating a single Fe bubble, and to explore how the 
presence of multiple Fe bubbles changes the global dynamics of young SNRs.

\section{HYDRODYNAMIC SIMULATIONS}
 
We use a parallel version (implemented with Message Passing Interface) of 
the Virginia Hydrodynamics (VH-1) numerical code 
to study the complex dynamics of SNRs with Fe bubbles. The simulations 
were computed on a spherical grid of $400^3$ zones covering an angular
span of 1.6 steradians.  This resolution is comparable to the simulations
by Chevalier at al. (1992) that used a grid of $256^2$.  
A higher resolution simulation
would show more small scale mixing, but the large-scale dynamics
of the problem would not change \cite{cbe92}.
Periodic boundary conditions were applied in both
the $\theta$ and $\phi$ directions.  
This choice for $\theta$ is somewhat unorthodox, but we wished to avoid the 
numerical artifacts associated with the coordinate
singularity at $\theta=0$, so we centered our grid about the 
equator ($0.3\pi < \theta < 0.7\pi$).  We experimented with
other boundary conditions (e.g., 
reflecting or zero gradients), but they proved unsatisfactory for various
reasons.  The numerical grid was expanded to follow the forward blastwave so the 
evolution could be tracked for many expansion times.

The simulations were initialized with a 
self-similar driven wave (SSDW) solution \cite{chevalier82} with an ejecta
density power law of $n=9$ expanding into either a uniform ambient density
(an ambient density power law of $s=0$) or a relic stellar wind (an ambient 
density power law of $s=2$).  The radial boundary conditions were set 
to match these power laws.
This self-similar structure 
has the following components: an outer blast wave, a wavy (dynamically 
unstable)
contact discontinuity between the shocked ambient medium and the shocked SN 
ejecta, and an inner reverse shock.  For these parameters, the forward shock
decelerates according to $R_s\propto t^{6/7}$ ($t^{6/9}$ for $s=0$), and 
the reverse shock
is located at a radius of $0.785 R_s$ ($0.842 R_s$ for $s=0$).  The initial conditions are spherically 
symmetric, so the narrow shell of shocked ejecta starts off smooth
but eventually shows signs of convective instability \cite{cbe92}.

\section{DYNAMICS WITH A SINGLE BUBBLE}

\begin{figure}[!hbtp]
\centerline{\includegraphics[width=16cm]{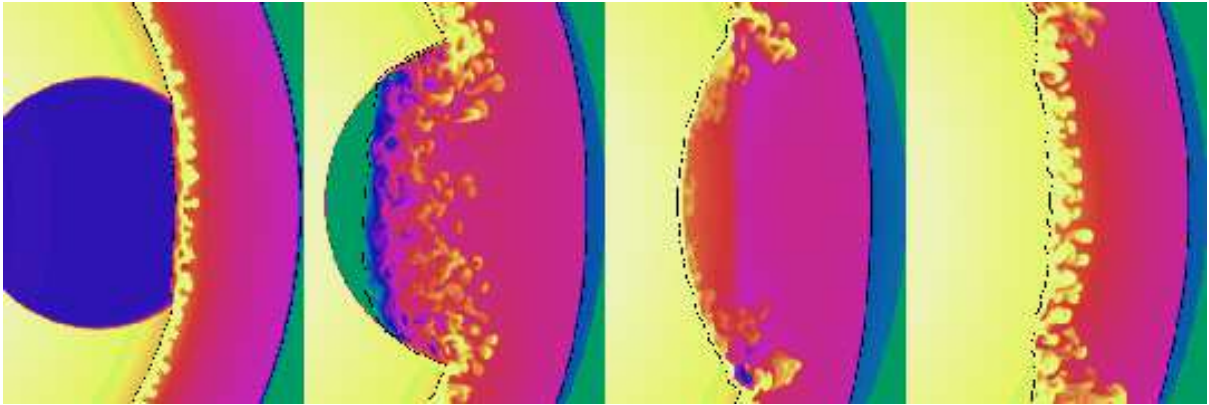}}
\caption{Dynamical interaction of an Fe bubble with a SSDW.  These 2D slices
through the center of the bubble are colored according to
gas density, with a logarithmic scale spanning a
range of 3.5 orders of magnitude (yellow=high, blue=low). The black lines
mark the surface of the forward and reverse shocks.  The thin blue strip
on the right of each panel marks the undisturbed ambient gas between the
forward shock and the edge of the numerical grid.  Note the slight deviation
of the shock from spherical.}
\label{fig:single}
\end{figure}

Hydrodynamical simulations are shown in 
Figure \ref{fig:single} of a SSDW ($n=9$, $s=2$)
with a single Fe bubble in the supersonic ejecta. 
The bubble, with a fractional radius of 0.4 (the radius of the bubble is
0.4 times the distance of the center of the bubble from the center of
the SN explosion), is 100 times less dense than 
the ambient SN ejecta at a radius corresponding to the center of the bubble. 
The pressure inside the bubble is equal to the pressure of the surrounding
ejecta in order to minimize motions, but the Mach number of expansion is
still relatively high within the bubble.  

The initial contact between the Fe bubble and the reverse shock is
illustrated in Figure \ref{fig:single} (left). When the outer edge of
the bubble reaches the reverse shock, the reverse shock accelerates into
the bubble because of its low density, while a rarefaction
wave propagates back into the shocked ejecta. 
The pressure behind the reverse shock
propagating through the bubble drops to more than an order of magnitude
smaller than the nominal pressure between the forward and reverse
shocks in the spherical SSDW.  There is only a modest drop in pressure
(compared to the spherical SSWD) in the shocked ambient medium ahead of the 
bubble because the SNR expansion timescale is comparable
to the sound crossing time across the bubble.
The high pressure of the relatively tenuous ambient gas then drives 
dense shocked ejecta deep into the bubble's 
interior. This situation is Rayleigh-Taylor
unstable, which leads to vigorous mixing of dense shocked ejecta with
the low-density shocked Fe from the bubble and the shocked ambient
gas.  This enhanced mixing is evident in the middle-left frame of
Figure \ref{fig:single}.  Note that although the reverse shock is moving
inwards in this expanding frame, at no time in the simulation is gas
ever moving radially inwards in the fixed frame of the explosion.

Eventually the
reverse shock reflects off the bottom of the bubble (middle-right panel
of Figure \ref{fig:single}) and propagates back through the layer of
mixed gas with a shock Mach number of $\sim 1.5$. 
The reverse shock is also transmitted into 
the dense ejecta, reforming the thin layer of shocked ejecta present in
the initial SSDW.  The bottom edge of the bubble is driven back out
to the original SSDW by the high density of the ejecta, and in the
process the extended layer of mixed gas inside the bubble
is swept up and compressed into a thin layer.

The shock reflected from the bubble's bottom eventually overtakes the blast
wave and the layer of shocked ejecta is pushed back to the original
radius of the reverse shock.   This marks the transition back to the original
self-similar stage, and the end of the transient stage associated with the 
bubble's presence.

\begin{figure}[!hbtp]
\centerline{\includegraphics[height=8cm]{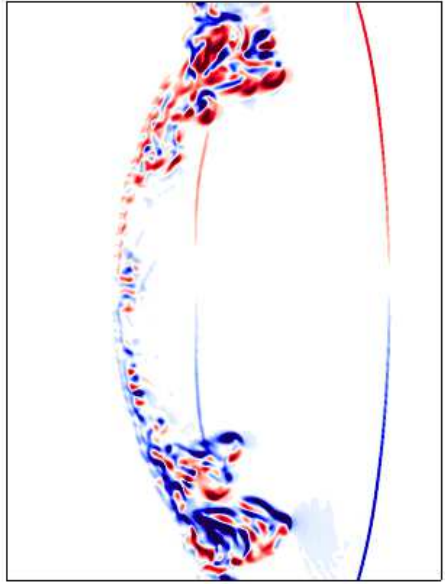}}
\caption{A 2D slice through the single bubble simulation, showing the
strong vorticity generated in a torus around the side of the bubble.
The color corresponds to vorticity (red=positive, blue=negative, white=0).
The slightly oblique forward and reflected reverse shocks show up as
thin lines of red and blue.  This time slice is the same as the middle-right
panel in Figure \ref{fig:single}.}
\label{fig:vort}
\end{figure}

What is less apparent, but perhaps most important, in this simulation is
the role of the oblique shocks along the bubble walls in generating 
vorticity.  As the reverse shock traverses the low-density bubble,
it drives an oblique shock into the wall of dense ejecta surrounding
the bubble.  This oblique shock 
generates vorticity at the bubble's boundary, and the thin layer
of shocked ejecta 
along the bubble's wall quickly becomes unstable.  This effect is
most pronounced along the sides of the bubble where the transmitted
shock is very oblique.  
By the time the
bubble has been completely shocked, this excess vorticity has piled
up in a thick ring with a radius slightly smaller than the original
Fe bubble, as shown in Figure \ref{fig:vort}.  It is this torus of 
high vorticity where the most vigorous
mixing occurs.  
This mixing can lead to relatively large velocities, with some shocked
ejecta traveling outwards 50\% faster than the gas immediately behind
the forward shock.

A hydrodynamical problem of a plane shock interacting with a low density
spherical (or cylindrical) bubble has been of interest to the fluid
dynamics community. Experiments (e.~g., Haas \& Sturtevant 1987) and
hydrodynamical simulations (e.~g., Quirk \& Karni 1996) revealed how 
vortex lines (for cylindrical bubbles) and vortex rings (for spherical
bubbles) form in this interaction process, although a complete understanding
of this complex interaction (including details of mixing of bubble's material 
with the ambient medium) is still lacking. Although geometry and dynamics
in our simulations are more complex than in these idealized experiments and
simulations, the emergence of vortex rings in our simulations is in 
qualitative agreement with
results reported in the literature. We refer the reader to a recent
review by Zabusky (1999) for further details about generation of vorticity in 
shock-accelerated inhomogeneous flows. 

This hydrodynamical problem of a shock interacting with a low density bubble
can be compared with the more often studied problem of a shock
interacting with an overdense cloud (e.~g., Klein et al. 1994; Xu \& 
Stone 1995; see also Zabusky 1999 and references therein).  In
both cases the important long-term result of the shock interaction is
the generation of vorticity and a rapid mixing of the gas inside and
outside the bubble/cloud. This is demonstrated in a dramatic fashion below
when we consider SN ejecta with multiple bubbles instead of a single bubble.

\section{DYNAMICS WITH MULTIPLE BUBBLES}

\begin{figure}[!hbtp]
\centerline{\includegraphics[width=16cm]{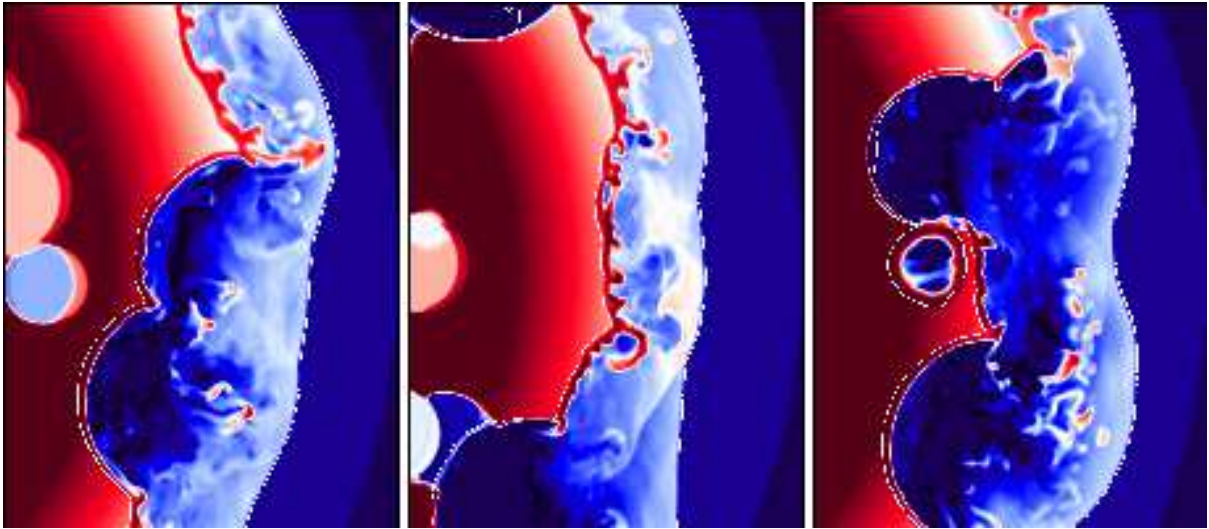}}
\caption{Dynamical interaction of multiple Fe bubbles with a SSDW
propagating through a uniform ambient medium ($s=0$).
These 2D slices through the center of the simulation show the gas
density (red=high, blue=low) along with pressure contours marking
the location of the forward and reverse shocks.  }
\label{fig:s0slices}
\end{figure}

\begin{figure}[!hbtp]
\centerline{\includegraphics[width=16cm]{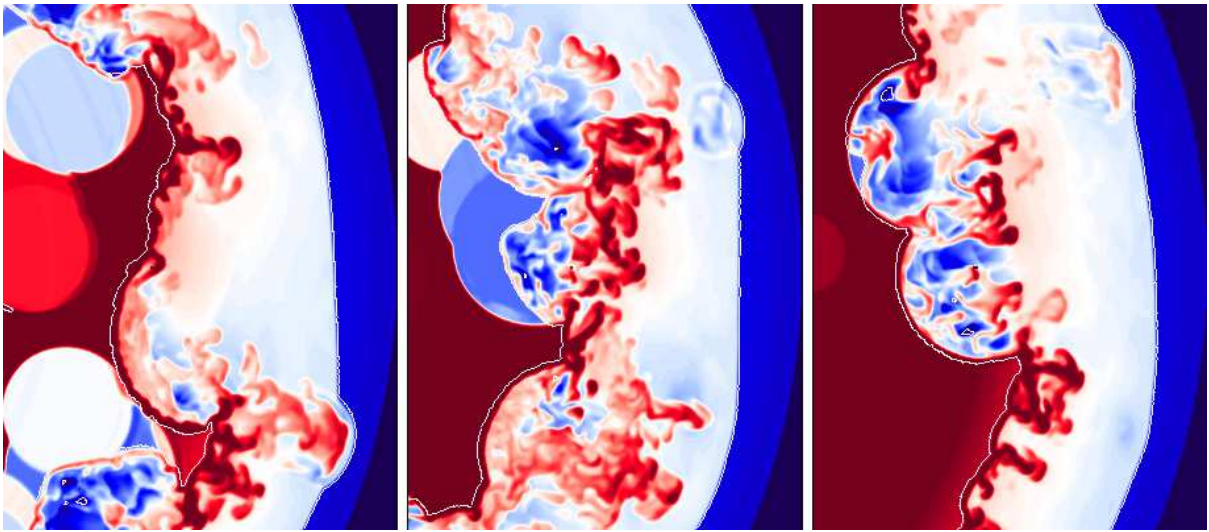}}
\caption{Dynamical interaction of multiple Fe bubbles with a SSDW
propagating through a relic wind ($s=2$).
These 2D slices through the center of the simulation show the gas
density (red=high, blue=low) along with pressure contours marking
the location of the forward and reverse shocks.  Note the relatively
common protrusions through the forward
shock driven by clumps of shocked ejecta.}
\label{fig:s2slices}
\end{figure}

A more realistic model of Fe bubbles in SN ejecta should include multiple
bubbles occupying a substantial fraction of the ejecta volume.  We have 
evolved such models for both $s=0$ (Figure \ref{fig:s0slices}) and
$s=2$ (Figure \ref{fig:s2slices}).
In these examples, bubbles with a fractional radius
of 0.2 and with a uniform density 100 times smaller than the ejecta density
are distributed randomly in the ejecta, with an average volume filling
fraction of $\sim 0.5$.  We tracked the filling fraction by computing the
fractional area occupied by bubbles at a radius of $0.5 R_s$.
Due to the clustering of bubbles created by the random number generator,
the filling fraction went as high as 0.7 and as low as 0.3, but averaged
reasonably close to 0.5.

\begin{figure}[!hbtp]
\centerline{\includegraphics{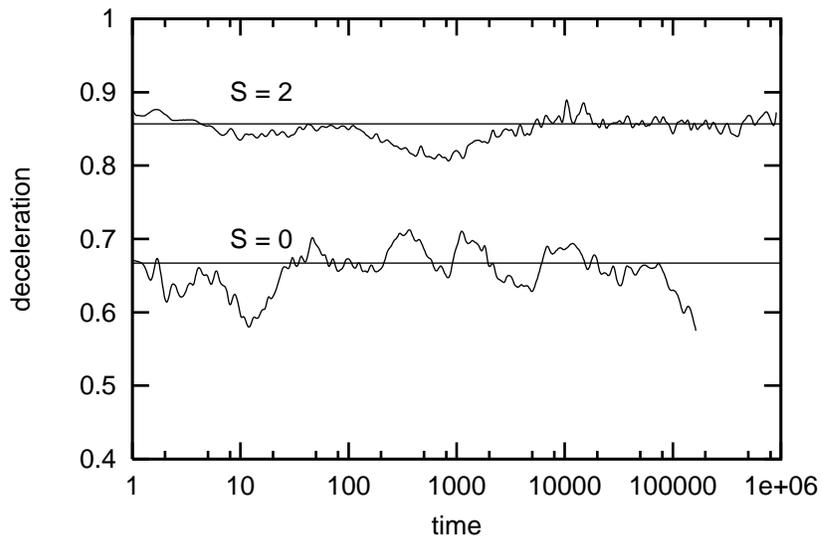}}
\caption{Time evolution of the deceleration parameter ($V_s t/R_s$)
for the simulations with multiple bubbles.  The straight lines 
correspond to the analytic values of the deceleration parameter
for the corresponding spherically symmetric SSDW.}
\label{fig:vtr}
\end{figure}

\begin{figure}[!hbtp]
\centerline{\includegraphics{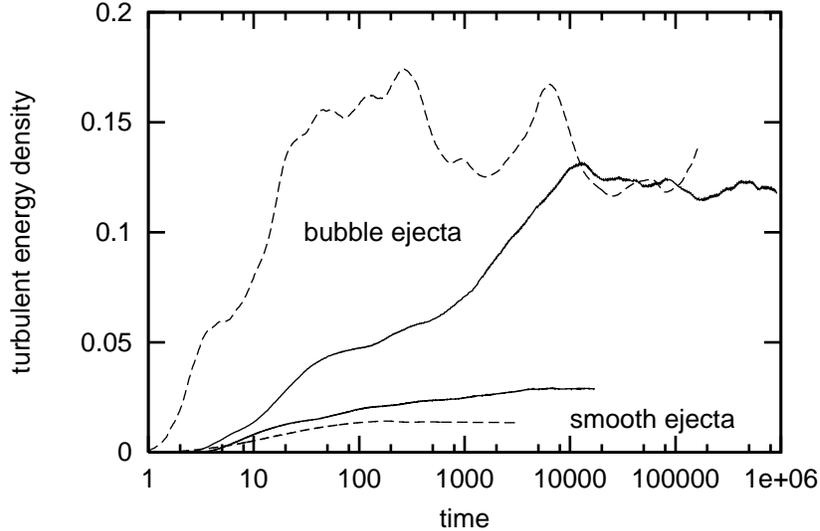}}
\caption{Time evolution of the turbulent energy density within the intershock region,
for the  simulations with multiple bubbles.  Dashed lines:  $s = 0$.  Solid
lines:  $s=2$.}
\label{fig:evolve}
\end{figure}

The outer shock wave
is relatively unaffected by the presence of the Fe bubbles in these
simulations; the shape of the outer shock remains relatively spherical
and the deceleration parameter ($Vt/R_s$) is within a few percent of the 
analytic solution of 6/7 for a spherical SSDW with $n=9$ and $s=2$
(6/9 for $s=0$).  The time evolution of the deceleration parameter 
from both runs is shown in Figure \ref{fig:vtr}.  The
SSDW has been evolved more than six orders of
magnitude in radius, requiring 30,000 timesteps, in order to allow the
blastwave to reach a quasi self-similar form (see the time-dependence
of turbulent energy density in Figure \ref{fig:evolve}).

In contrast to the forward shock, the presence of multiple bubbles
dramatically alters the reverse shock and makes the interaction region
very turbulent and inhomogeneous.  From the results of the single bubble
simulation, we expect the position of the reverse shock to vary randomly on 
a scale of order the radius of the bubbles.  This variation is illustrated
in the 2D slices of the multiple bubble simulation shown in 
Figures \ref{fig:s0slices} and \ref{fig:s2slices}.
The deviations can in fact be much larger than a bubble diameter because
of the overlap of bubbles when the filling factor is large.  In the $s=2$ model
the reverse shock came close to the inner radial boundary at $0.5R_s$
at several times during the simulation.  In the $s=0$ model we were forced
to extend the radial grid beyond $0.5R_s$ in order to keep the reverse 
shock contained within the simulation domain.  
In addition to a strong variation in the radius of the reverse shock, the
average radius of the reverse shock is significantly smaller than in the
SSDW solution. The average 
width of the interaction
region grows by $\sim 50\%$ during the first half of the simulations, 
from the analytic value of $0.21$ to an average
value of $0.31$ for $s=2$ and from $0.16$ to $0.26$ for $s=0$.  

\begin{figure}[!hbtp]
\centerline{\includegraphics{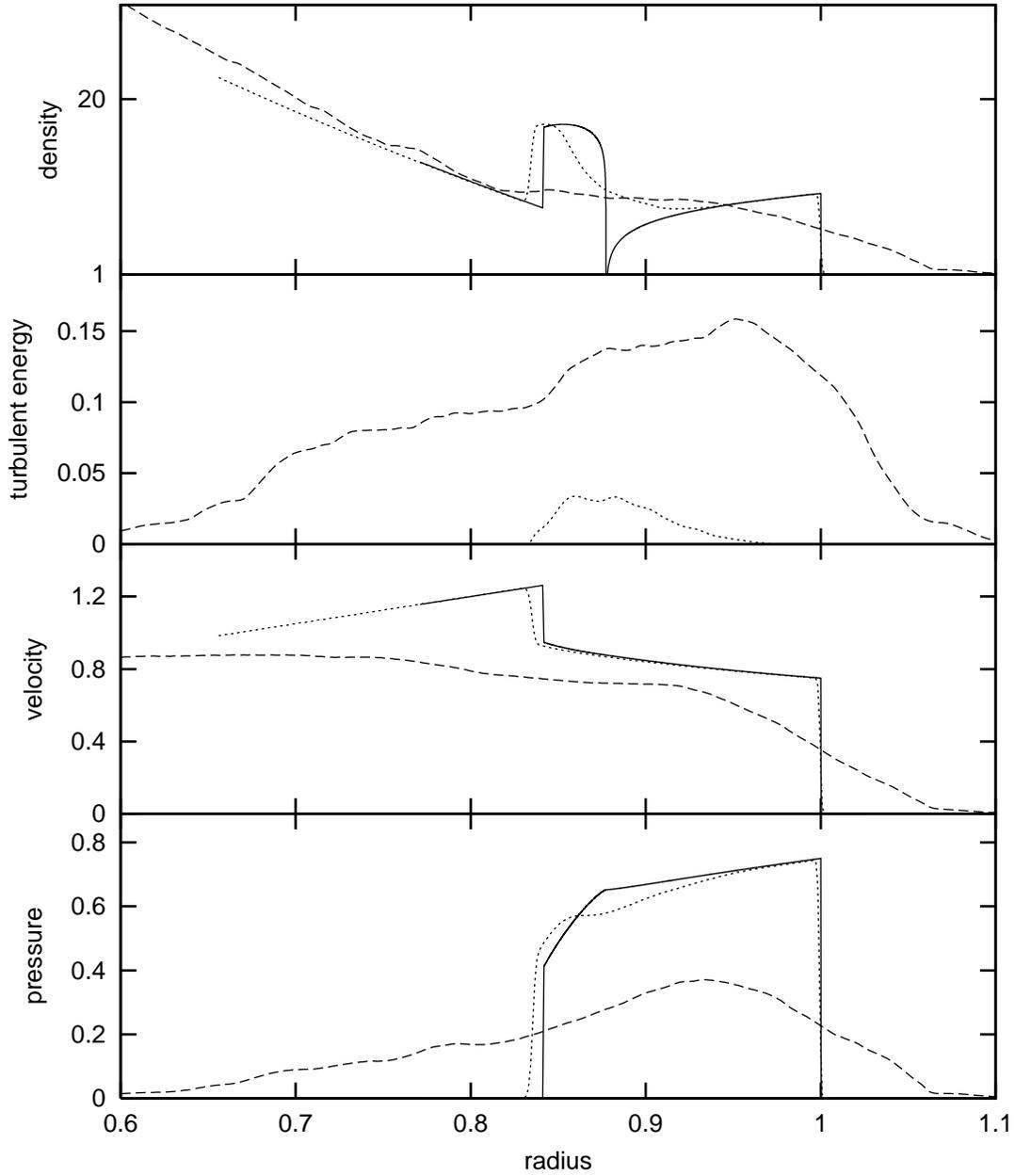}}
\caption{Angle-averaged radial profiles for $s=0$, comparing the
3D Fe-Ni bubble simulation (solid lines), a 3D simulation with 
smooth ejecta (dashed lines), and the analytic 1D solution (dotted lines).
The density is scaled to the preshock density, $\rho_0$, the velocity
is scaled to the shock velocity, $V_s$, and the pressure and turbulent
kinetic energy density are scaled to $\rho_0 V_s^2$.}
\label{fig:profile0}
\end{figure}

\begin{figure}[!hbtp]
\centerline{\includegraphics{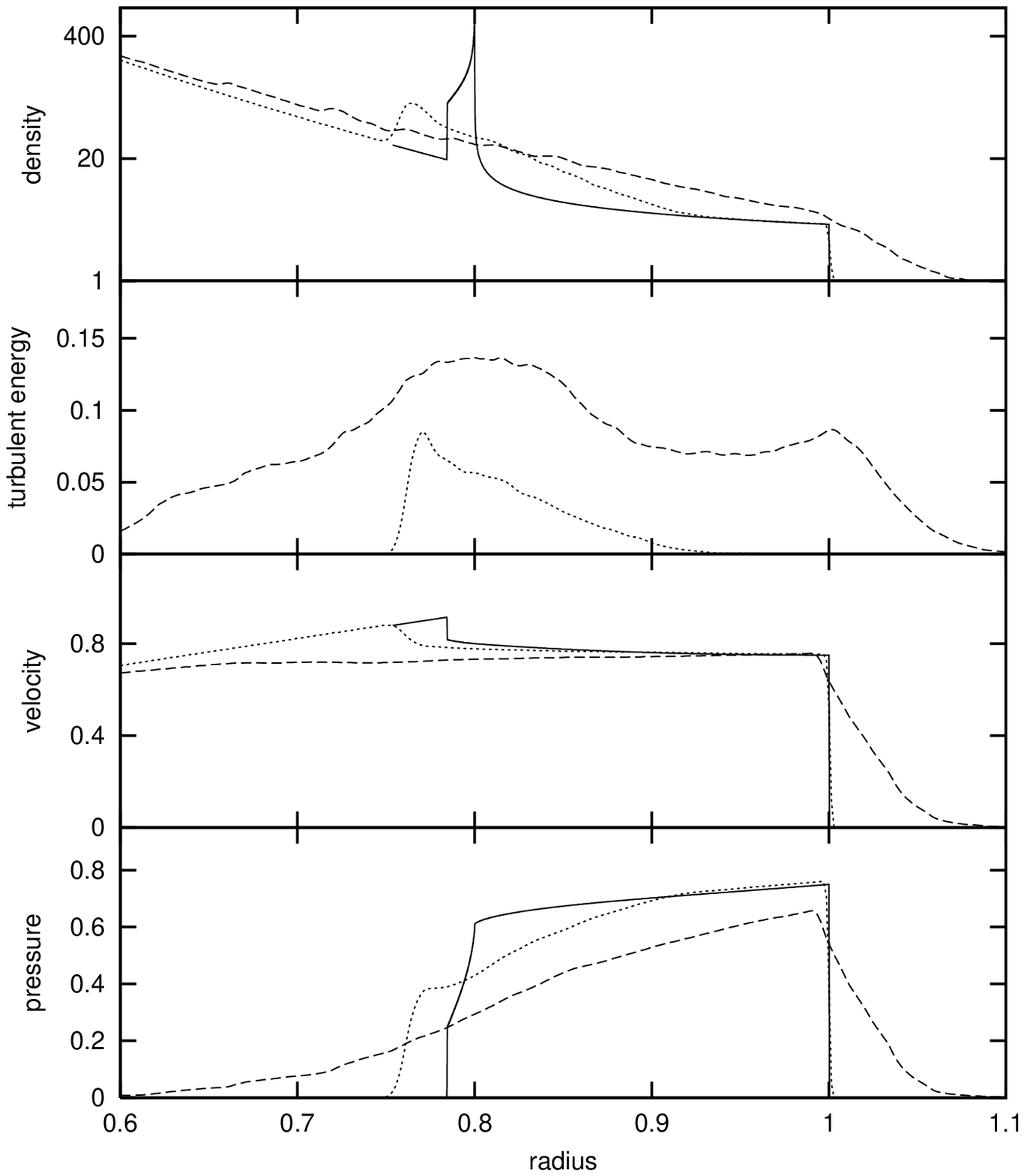}}
\caption{Angle-averaged radial profiles for $s=2$, comparing the
3D Fe-Ni bubble simulation (solid lines), a 3D simulation with 
smooth ejecta (dashed lines), and the analytic 1D solution (dotted lines).
The density is scaled to the preshock density, $\rho_0$, the velocity
is scaled to the shock velocity, $V_s$, and the pressure and turbulent
kinetic energy density are scaled to $\rho_0 V_s^2$.}
\label{fig:profile2}
\end{figure}

Angle-averaged radial profiles of density, turbulent energy, velocity,
and pressure for $s = 0$ (Fig. \ref{fig:profile0}) and $s = 2$
(Fig. \ref{fig:profile2}) quantitatively demonstrate how the presence of
Fe bubbles affects the SNR structure. Density and velocity 
profiles are remarkably smooth when compared with analytic and numerical
solutions for homogeneous ejecta. A spatially-distinct shell of shocked
ejecta visible in these solutions is no longer discernible in simulations
with bubbles. This nearly complete obliteration of radial structure is
not surprising in view of the irregular shape of the reverse shock just
discussed. The increase
in the width of the interaction region mentioned above is most clearly 
demonstrated by the broad pressure profiles.
Angle-averaged pressures are generally lower than for homogeneous ejecta
because of the presence of unshocked ejecta with negligible pressure
throughout the interaction region. In the vicinity of the blast wave 
angle-averaged pressure is also lower because of the presence of the 
unshocked ambient gas located between the blast wave protrusions.

The irregular shape of the reverse shock has an important effect on the
dynamics of the SNR.  If a region of interbubble ejecta
encounters a relatively oblique reverse shock, it will be decelerated
less than it would in a spherically symmetric SSDW.  This dense region
would then propagate through the intershock region to produce a
protrusion of the forward shock.  In effect, the geometry of the
interbubble ejecta produced small overdense regions that behaved more
like clumps within low-density ejecta than bubbles within 
high-density ejecta.  The regions of interbubble ejecta that appeared
to have the most effect on the intershock region were comparable in
size to the bubbles themselves.  Larger regions of ejecta without
bubbles resulted in a relatively planar reverse shock, while smaller
regions of ejecta between bubbles did not significantly perturb the
intershock region.

However, even in the most favorable scenario, regions of dense shocked
ejecta traveling through the intershock region were quickly disrupted
by fluid instabilities and ultimately had a minimal impact on the
forward shock.  Dense clumps being slowed by the more tenuous shocked
ambient gas are Rayleigh-Taylor unstable, while shear flow around the
clumps creates Kelvin-Helmholtz instability.  Both processes act to
spread out the clump laterally and increase the drag.  The result is
that any dense clump creating a protrusion of the forward shock is
almost immediately sheared apart, with the remaining pieces quickly
advected back into the intershock region.  Note that these same
processes are responsible for shaping the shell of shocked ejecta in
the case of spherically-symmetric ejecta \cite{cbe92}.  In the absence
of Fe bubbles, however, the drag within the intershock region
prevented the shocked ejecta from getting anywhere close to the shock
front.

\begin{figure}[!hbtp]
\centerline{\includegraphics[width=16cm]{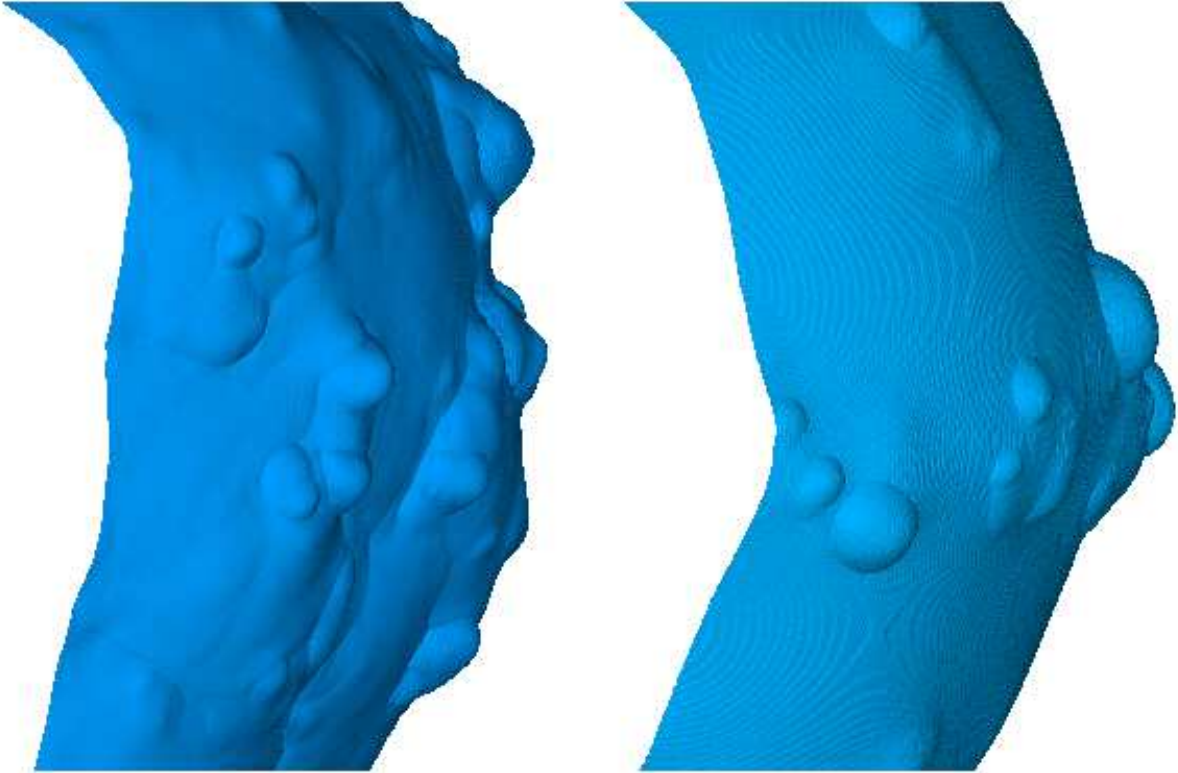}}
\caption{Deformations of the forward shock due to Fe bubbles
in the SN ejecta, for $s=0$ (left) and $s=2$ (right).}
\label{fig:surface}
\end{figure}

The impact of these ejecta clumps on the shape of the forward shock
is shown in Figure \ref{fig:surface}.  The
most pronounced deformations of the forward shock were relatively
small wavelength (less than a bubble radius) protrusions driven by
clumps of shocked ejecta.  Some of these protrusions pushed the forward
shock out an extra 15\%, while longer wavelength variations in the 
radius of the forward shock were limited to only a few percent. Angle-averaged
profiles (Figs. \ref{fig:profile0} and \ref{fig:profile2}) show that 
$\le 5$\%\ excursions of the blast wave are common.

Wang \& Chevalier (2000) arrived at similar conclusions regarding clumpy
ejecta, namely that even extremely overdense clouds have relatively little
effect on the forward shock.  They used 2D simulations to study the
evolution of a small, dense cloud placed in the ejecta of a Type Ia SN
model in which the nominal ejecta are described by an exponential density
profile.  As long as sufficient spatial resolution was provided, their
overdense clumps were flattened out and broken up by fluid instabilities
before they reached the shock front.  In 3D and with higher numerical
resolution, this process should happen even faster.

The dynamics in these multiple-bubble simulations appear much richer
than isolated bubbles or clouds alone.
In the multiple bubble simulations the propagation of the reverse shock
through bubbles and around interbubble regions
leads to
a substantial amount of turbulence in the intershock region.  We tracked
the turbulent energy density in the intershock region by summing up
the transverse (non-radial) kinetic energy in each zone between the
reverse and forward shocks.  We adjusted this value by $3/2$ to account
for the radial component of the turbulent flow (i.e., we assume all three
components of the turbulent flow have comparable magnitude: 
$v_r^2 \sim v_\theta^2 \sim v_\phi^2$).  This average was then normalized
by the kinetic energy density flowing through the shock front, $\rho_o V_s^2$.
The results are plotted as a function of radius in Figures
\ref{fig:profile0} and \ref{fig:profile2}, and as a function of time in 
Figure \ref{fig:evolve}. In contrast to the case of homogeneous ejecta, the 
turbulent region is no longer restricted to the shocked ejecta and the
adjacent shocked ambient gas. The turbulence extends across the whole
interaction region, as implied by broad, slowly varying turbulent energy 
profiles. The average turbulent
energy density grows relatively slowly, but eventually reaches a significant
fraction ($>10\%$) of the energy density associated with the forward shock.

\section{OBSERVATIONAL CONSEQUENCES}

\subsection{Gas Temperature and Velocity}

The irregular shape of the reverse shock has important implications
regarding the X-ray spectrum produced by these SNR models.
Any deviation from a spherical shock will decrease the shock velocity and hence
the postshock temperature.  From Figures \ref{fig:s0slices} 
and \ref{fig:s2slices} we see that this
will be mildly important for the forward shock, but can be expected to
have dramatic effects for the reverse shock.  Furthermore, as the reverse
shock moves backward through a bubble (in the expanding frame), the shock
velocity can be higher than in the spherical case, resulting in higher
postshock temperatures (but with small emission measure because of the
low density).  

To estimate these effects without undertaking the complicated effort of
calculating X-ray spectra, we have summed up the emission measure (EM) for
each zone in the simulation, and plotted this as a function of
temperature and radial velocity (scaled to the postshock temperature
and shock velocity). (This is only a very approximate procedure as we 
neglected variations in the mean molecular weight between Fe bubbles and
the ambient ejecta. For detailed comparisons with X-ray observations one 
also needs to consider electron temperature which is generally lower than
the mean gas temperature discussed here.)

\begin{figure}[!hbtp]
\includegraphics[width=16cm]{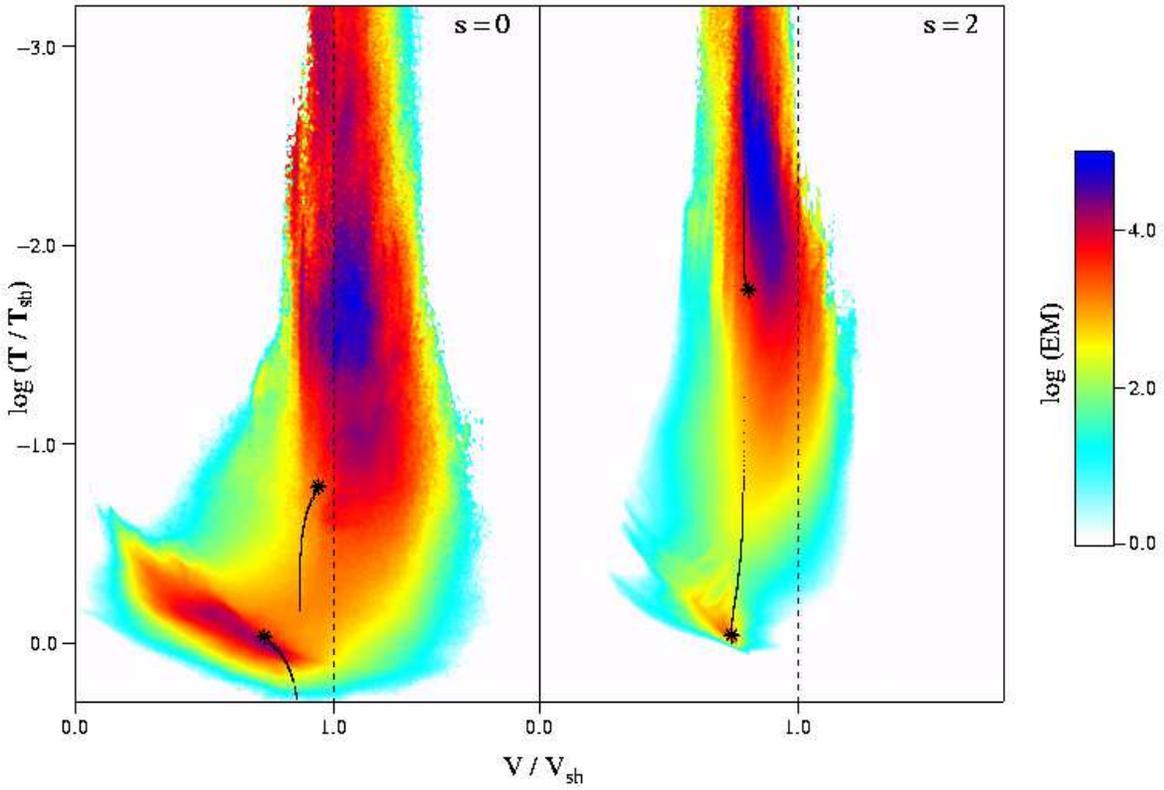}
\caption{A map of the emission measure as a function
of temperature and radial velocity for both the $s=0$ and
$s=2$ simulations.  The temperature is scaled to the postshock temperature,
and the velocity is scaled to the velocity of the forward shock.  
The black lines map out the region of emission for
the corresponding spherical SSDW, with stars at their ends marking shock 
locations. Note that while the analytic solutions
differ considerably between $s=0$ and $s=2$, the results of the 3D hydrodynamic
simulations with Fe bubbles are quite similar.}
\label{fig:emission}
\end{figure}

\begin{figure}[!hbtp]
\includegraphics[width=16cm]{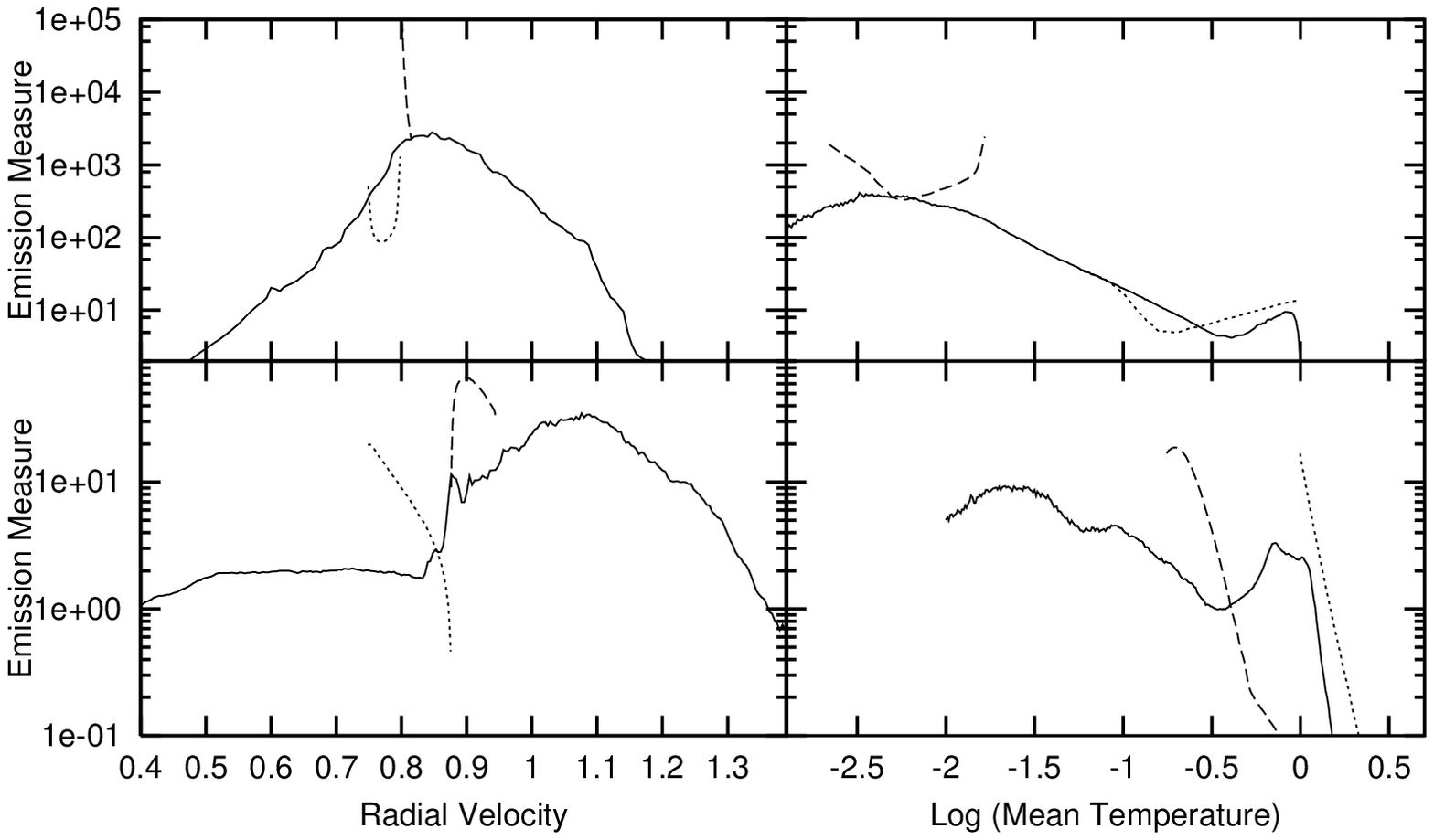}
\caption{The emission measure as a function
of temperature and radial velocity for both the $s=0$ (bottom) and
$s=2$ (top) simulations.  The temperature is scaled to the postshock temperature,
and the velocity is scaled to the velocity of the forward shock.  
The solid lines correspond to the 3D simulation with Fe-Ni bubbles.  
For comparison, we also plot the emission measure from the shocked
ejecta (dashed lines) and shocked ambient medium (dotted lines) in
the corresponding 1D self-similar solutions.}
\label{fig:emissionlines}
\end{figure}

The results for both the $s=2$ and $s=0$
multiple bubble simulations are shown in Figures
\ref{fig:emission} and \ref{fig:emissionlines}.
Superimposed on these plots are the curves corresponding to emission
from spherically symmetric SSDWs.  In both cases these curves are
composed of two pieces, one for the shocked ejecta and one for the
shocked ambient gas.  Shocked gas immediately behind the forward shock
is located at $\log (T) = 0$ and $V = 0.75$.  For the $s=2$ case the
gas temperature in the SSDW decreases with increasing distance from the
forward shock, while the opposite holds for $s=0$.  Shocked ejecta
immediately behind the reverse shock are located at $\log (T) = -1.76$
and $V = 0.82$ and decrease in temperature away from the shock in the
case of $s=2$.  For $s=0$, the shocked ejecta start off at $\log (T) =
-0.75$ and $V = 0.95$ and increase in temperature away from the
shock.  
Despite these differences in the spherically symmetric models,
the emission maps from the 3D simulations with multiple bubbles appear
relatively similar.  In both cases X-ray emitting gas is significantly
spread out in temperature.  This is most dramatic in
the $s=0$ case, for which there is no cool gas in the spherical
solution.

There is also a large
spread in the radial velocity of the X-ray emitting gas, particularly 
for the case of $s=0$.  In both simulations the radial velocity of
the shocked ambient gas is spread to lower velocities as a result of
protrusions in the forward shock, while the radial velocity of the 
shocked ejecta is spread to higher velocities.  In the case of $s=0$,
the bulk of the X-ray emitting ejecta is traveling faster than the
forward shock.
\subsection{SNR Morphology}

To provide a qualitative estimate of how Fe bubbles might affect the
observed morphology of SNRs, we created a volume rendering of the
emission measure of shocked gas computed from the last frame in our
multiple-bubble simulations.  While the actual emission observed from
a SNR will also depend on the local temperature, abundance, and ionization,
the emission measure provides a simple, convenient means for exploring
the X-ray morphology implied by a given hydrodynamical model.  As such, these
images are meant only to show an overall  qualitative agreement with 
the observed X-ray morphology of SNRs like Cas A.

\begin{figure}[!hbtp]
\includegraphics[width=16cm]{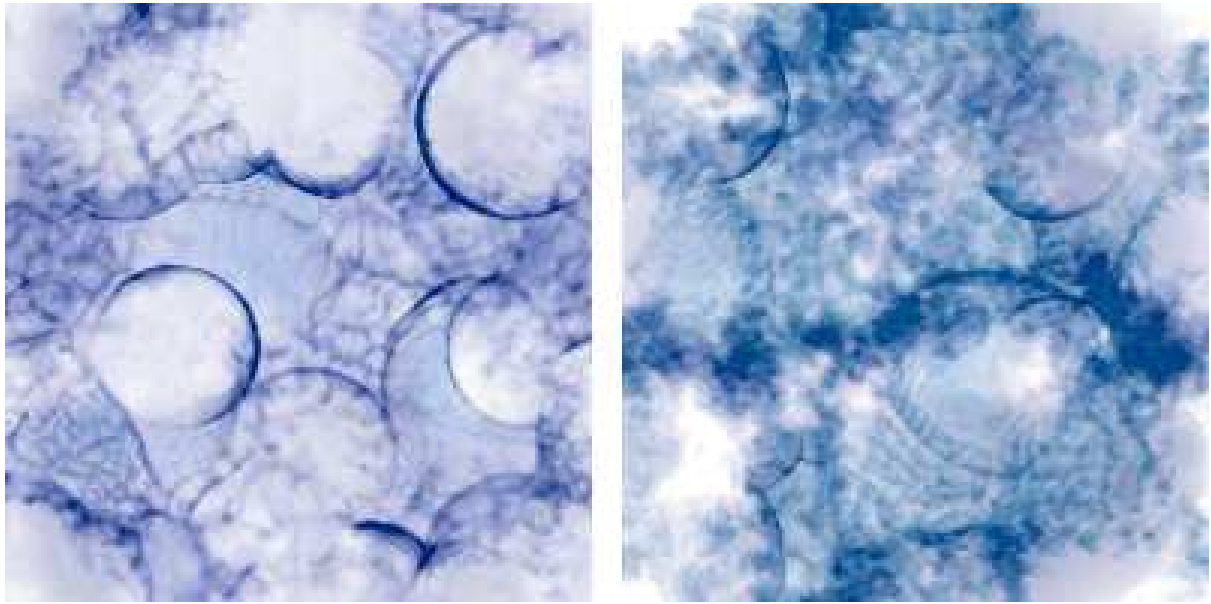}
\caption{A volume rendering of the gas density from the 
multiple bubble simulations ($s=0$ on the left, $s=2$ on the right), 
illustrating the 
prominence of the filaments of shocked ejecta.  This view is looking down
onto the simulation such that the SSDW is propagating up out of the page.}
\label{fig:render}
\end{figure}

The rendered images shown in Figure \ref{fig:render} are dominated by
narrow filaments of shocked ejecta.  These filaments are barely resolved in 
our numerical simulations, and they occupy a very small fraction of the
volume of the SNR.  In the $s=2$ model, 90\% of the EM comes from less
than 2\% of the volume of shocked gas.  These filaments often show up
as rings or partial rings representing the circumference of Fe bubbles
as they pass through the reverse shock.  These rings show up more prominently
in the $s=0$ simulation.

\section{SUMMARY}

Our primary conclusion is that large-scale inhomogeneities associated with 
the presence of Fe bubbles in heavy-element ejecta lead to vigorous 
turbulence and mixing in young SNRs. The turbulent energy density may be 
increased by an order of magnitude when compared with homogeneous ejecta.
This turbulence radically changes the spatial structure of the interaction
region between the blast wave and the reverse shock.
While the time-averaged dynamics and
the propagation of the forward shock are well described by a self-similar
solution \cite{chevalier82,cbe92}, the reverse shock geometry and the
structure of the shocked region are very different in the presence of the
Fe bubbles. The reverse shock is no longer approximately spherically
symmetric, and jet-like fingers of ejecta can even affect the location of
the blast wave. Angle-averaged density profiles rise smoothly toward the
remnant's interior, with no signs for the presence of a spatially distinct 
shell of ejecta.  Pressure and temperature variations of orders of
magnitude are 
also present, unlike the case of homogeneous ejecta which can be described by
well-defined density, pressure, and temperature profiles. Such large 
variations should dramatically influence X-ray emission, which depends strongly
on temperature and density. In particular, low temperature, dense ejecta
are particularly efficient in producing X-ray emission, and may dominate
X-ray spectra. But this emission is generally produced by ejecta which
suffered the least amount of deceleration, so that their radial velocities
and proper motions may be even higher than that of the blast wave. This
example demonstrates that standard SSDW solutions should be used with an
extreme caution for young SNRs with inhomogeneous ejecta.
The amount of turbulence generated by the presence of Fe bubbles
in SNRs most likely varies from remnant to remnant because of large 
($\sim 100$) variations in the radioactive Ni yields in core-collapse SNe. 
In addition, the amount of turbulence might be lower for those SNRs which
did not have enough time to enter the self-similar turbulent regime, 
because of long timescales necessary to achieve it.
But unless mixing of heavy elements is somehow inhibited during
a SN explosion, it should be clear that one-dimensional models are clearly 
not acceptable for remnants of core-collapse SNe, and that
multidimensional hydrodynamical modeling is essential for understanding
dynamics of heavy-element ejecta in such remnants.  It is also
possible that Ni bubbles can be formed in Type Ia supernova
explosions (Wang \& Chevalier 2000), so that our results
may be applicable there as well.

We expect that small-scale turbulence will have a strong effect on the
synchrotron radio morphology of a young remnant.  In 2-D MHD
simulations of the SSDW phase of a young Type Ia ($s = 0$) supernova
remnant, Jun \& Norman (1996) found that turbulent energy eventually
rose to about 0.6\% of the kinetic energy in the remnant, and that
random magnetic energy rose as well, to a level of about 0.3\% of the
turbulent energy.  The random magnetic-field energy did not track the
total turbulent energy precisely, but spatial locations of strong
turbulence (Kelvin-Helmholtz unstable edges of Rayleigh-Taylor
fingers) were also the locations of strong magnetic-field
amplification.  In synchrotron visualizations, Jun \& Norman (1996)
also found that regions of strong magnetic field produced strong
synchrotron emission, but their simulation used a relatively simple
description of relativistic-electron acceleration and transport.  We
expect that magnetic fields in a bubble simulation would also track
locations of turbulence and vorticity, that is, should indicate bubble
walls, and ought to be noticeable in synchrotron images.  In addition,
the far higher levels of turbulent energy we find compared to smooth
ejecta models such as Jun and Norman's should result in considerably
higher magnetic energy densities and consequent higher synchrotron
emissivities.  However, hydrodynamic simulations including
magnetic-field tracking and particle acceleration and subsequent
evolution will be necessary to make more definite predictions.

Our results are in qualitative agreement with the observed morphology of
Cas A, undoubtedly the best example of a young SNR with ample evidence
for vigorous mixing. Our simulations are able to reproduce filamentary
ejecta emission  seen in {\it Chandra} images. The ragged appearance of 
Cas A, including its ``jet'' feature, may also be a consequence of an
interaction of bubbly ejecta with the ambient medium, and not necessarily
the result of a strongly asymmetric explosion. It is also tempting to
identify its rings of optical knots (Reed et al. 1995) with the most dense
ejecta clumps at the boundaries of Fe bubbles. Our current simulations are 
however not suitable for a detailed modeling of the Cas A dynamics because 
of the evidence for the dynamically important shell of circumstellar matter 
in Cas A \cite{cl89}. The circumstellar interaction, studied by
us through one-dimensional hydrodynamical simulations \cite{bsbs96}, must 
be now simulated in 3-D in the framework of bubbly ejecta. A quantitative study
of spatial morphology of Cas A, coupled with a more detailed examination of
spatial structures in hydrodynamical data sets, would also be useful.

X-ray observations should provide a most complete test of the two-phase model,
because they probe the bulk of the
shocked material. While modeling of X-ray emission based on
multidimensional calculations clearly demands a separate effort, we briefly
outline what major effects are expected. If most
Fe initially resides in low-density shocked bubbles, while other heavy elements
are located in the dense phase of SN ejecta, then Fe lines should generally
be much weaker than lines from other abundant heavy elements. 
Fe lines indeed appear to be weaker than expected in 
Cas A \cite{bsbs96,vkb96,favata97} when compared with strong Si, S, Ar, 
and Ca lines, based on nucleosynthetic yields of core-collapse SNe. Unlike in
the optical and IR, freshly synthesized Fe was detected by {\it Chandra} 
\cite{hughes00, hwang00}, but its spatial distribution is different than that
of Si- and S-rich ejecta. This means that not all Fe resided in low-density
bubbles, either because of microscopic mixing of Fe at the bubbles' boundaries
or because radioactive Ni was mixed to sufficiently high velocities where
the bubble growth was inhibited by escape of $\gamma$-rays from Ni clumps.
Quantitative analysis of {\it Chandra} Cas A data, coupled with hydrodynamical 
simulations and X-ray modeling, should provide a stringent test of the 
two-phase ejecta model with Fe bubbles.

As remnants become older, turbulent mixing will lead to a gradual
strengthening of Fe lines with respect to other heavy elements. Even
if Fe is microscopically mixed with other elements during the SNR
evolution, its ionization age should lag behind that of other
elements. In the next few years, a combination of new X-ray
observations and sophisticated hydrodynamical and X-ray modeling
should provide us with a vastly better understanding of how chemical
elements are ejected in explosions of massive stars and how they are
dispersed into ambient interstellar medium.

\acknowledgements We thank Dick McCray for discussions about Fe-Ni bubbles
in SN 1987A. The three-dimensional simulations reported here were performed
at the North Carolina Supercomputing Center using 100 processors of
an IBM SP2.  We thank NCSC and IBM for their generous support of computing 
resources.
Support for this work was provided by NASA under grant NAG-7153.

\end{document}